\documentclass[twocolumn,english,aps,pra,showpacs,floatfix]{revtex4-1}
\usepackage[latin9]{inputenc}
\usepackage{amsmath}
\usepackage{amssymb}
\usepackage{wasysym}
\usepackage{graphicx}
\usepackage{esint}

\makeatletter
\@ifundefined{textcolor}{}
{%
 \definecolor{BLACK}{gray}{0}
 \definecolor{WHITE}{gray}{1}
 \definecolor{RED}{rgb}{1,0,0}
 \definecolor{GREEN}{rgb}{0,1,0}
 \definecolor{BLUE}{rgb}{0,0,1}
 \definecolor{CYAN}{cmyk}{1,0,0,0}
 \definecolor{MAGENTA}{cmyk}{0,1,0,0}
 \definecolor{YELLOW}{cmyk}{0,0,1,0}
}

\usepackage{babel}

\usepackage{babel}

\usepackage{babel}

\makeatother

\usepackage{babel}
\begin{document}

\title{Robust quantum gates for stochastic time-varying noise}

\author{Chia-Hsien Huang}

\author{Hsi-Sheng Goan}

\email{goan@phys.ntu.edu.tw}

\selectlanguage{english}%

\affiliation{Department of Physics and Center for Theoretical Sciences, National
Taiwan University, Taipei 10617, Taiwan}

\affiliation{Center for Quantum Science and Engineering, National Taiwan University,
Taipei 10617, Taiwan}

\date{\today}
\begin{abstract}
How to effectively construct robust quantum gates for time-varying
noise is a very important but still outstanding problem. Here we develop
a systematic method to find pulses for quantum gate operations robust
against both low- and high-frequency (comparable to the qubit transition
frequency) stochastic time-varying noise. Our approach, taking into
account the noise properties of quantum computing systems, can output
single smooth pulses in the presence of multi-sources of noise. Furthermore,
our method can be applied to different system models and noise models,
and will make essential steps toward constructing high-fidelity and
robust quantum gates for fault-tolerant quantum computation. Finally,
we discuss and compare the gate operation performance by our method
with that by the filter-transfer-function method. 
\end{abstract}


\maketitle
\maketitle
\maketitle
\maketitle
\renewcommand\figurename{FIG.}
\makeatletter
\renewcommand\thefigure{\@arabic\c@figure}
\renewcommand\fnum@figure{\figurename~\thefigure}  \newcommand\onlinecap{\renewcommand\fnum@figure{\figurename~\thefigure~(color online)}}
\makeatother

\section{Introduction}

To realize practical quantum computation, a set of high-fidelity universal
quantum gates robust against noise in the qubit system is prerequisite.
Constructing control pulses to operate quantum gates which meet this
requirement is an important and timely issue. Quantum gates in open
quantum systems have been investigated by various methods such as
dynamical decoupling methods \cite{KhodjastehViola2009,*KhodjastehViola2009a,*KhodjastehLidarViola2010,*KhodjastehBluhmViola2012,WestLidarFongEtAl2010,SouzaAlvarezSuter2012,FanchiniHornosNapolitano2007a,*FanchiniNapolitanoCiakmakEtAl2015,ChaudhryGong2012,XuWangDuanEtAl2012}
and optimal control methods \cite{ClausenBenskyKurizki2010,*ClausenBenskyKurizki2012,WeninPoetz2008,RebentrostSerbanSchulte-HerbrueggenEtAl2009,HwangGoan2012,*TaiLinGoan2014,*HuangGoan2014,*ChouHuangGoan2015}.
For classical noise, there are many robust control methods such as
composite pulses \cite{Tycko1983,Levitt1986,Wimperis1994,CumminsLlewellynJones2003,BrownHarrowChuang2004,VandersypenChuang2005,IchikawaBandoKondoEtAl2011,*BandoIchikawaKondoEtAl2013,IchikawaGuengoerdueBandoEtAl2013,MerrillBrown2014},
soft uniaxial positive control for orthogonal drift error (SUPCODE)
\cite{WangBishopKestnerEtAl2012,KestnerWangBishopEtAl2013,WangBishopBarnesEtAl2014,WangCalderon-VargasRanaEtAl2014,RongGengWangEtAl2014,YangWang2016},
sampling-based learning control method \cite{ChenDongLongEtAl2014,DongChenQiEtAl2015,DongWuChenEtAl2016},
inhomogeneous control methods \cite{LiKhaneja2006,LeeMontanoDeutschEtAl2013},
analytical method \cite{BarnesWangDasSarma2015}, single-shot pulse
method \cite{DaemsRuschhauptSugnyEtAl2013}, optimal control methods
\cite{KosutGraceBrif2013,AndersonSosa-MartinezRiofrioEtAl2015,GordonKurizkiLidar2008},
invariant-based inverse engineering method \cite{ruschhaupt2012optimally,LuChenRuschhauptEtAl2013},
and filter-transfer-function (FTF) methods \cite{GreenSastrawanUysEtAl2013,SoareBallHayesEtAl2014,BallBiercuk2015,Paz-SilvaViola2014}.
However, in most of these methods \cite{Tycko1983,Levitt1986,Wimperis1994,CumminsLlewellynJones2003,BrownHarrowChuang2004,VandersypenChuang2005,IchikawaBandoKondoEtAl2011,*BandoIchikawaKondoEtAl2013,IchikawaGuengoerdueBandoEtAl2013,MerrillBrown2014,WangBishopKestnerEtAl2012,KestnerWangBishopEtAl2013,WangBishopBarnesEtAl2014,WangCalderon-VargasRanaEtAl2014,RongGengWangEtAl2014,YangWang2016,ChenDongLongEtAl2014,DongChenQiEtAl2015,DongWuChenEtAl2016,LiKhaneja2006,LeeMontanoDeutschEtAl2013,BarnesWangDasSarma2015,DaemsRuschhauptSugnyEtAl2013,KosutGraceBrif2013,AndersonSosa-MartinezRiofrioEtAl2015},
noise is assumed to be quasi-static, i.e., is time-independent within
the gate operation time but can vary between different gates. We call
these robust control strategies the quasi-static-noise (QSN) methods.
But this QSN assumption is not always valid \cite{Oliver2014}. The
robust performance of control pulses obtained by the QSN methods under
time-dependent noise (e.g., $1/f^{\alpha}$ noise with $\alpha\apprge1$)
\cite{KabytayevGreenKhodjastehEtAl2014,WangBishopBarnesEtAl2014,WangCalderon-VargasRanaEtAl2014,YangWang2016}
have been investigated, and it was found that they can still work
well for relatively low-frequency non-Markovian noise.

Stochastic time-dependent noise is treated in the FTF method \cite{GreenSastrawanUysEtAl2013,SoareBallHayesEtAl2014,BallBiercuk2015}
in which the area of the filter-transfer function in the frequency
region, where the noise power spectral density (PSD) is non-negligible,
is minimized. However, in this approach only the the filter-transfer
function overlapping with the noise PSD in the preset frequency region
is considered, but the detailed information of the distribution of
the noise PSD is not included in the optimization cost function. Here
we develop an optimal control method in time domain by choosing the
ensemble average gate infidelity (error) as our cost function for
optimization. As a result, the noise correlation function (CF) or
equivalently the detailed noise PSD distribution appears naturally
in our chosen optimization cost function. Therefore our method can
have better robust performance against noise in a general case. The
idea of our method is simple, and our method is not limited to particular
system models, noise models, and noise CF's. We demonstrate our robust
control method for classical noise in this paper, but our method can
be easily generalized to the case of quantum noise by replacing the
ensemble average for classical noise with the trace over the degrees
of freedom of the quantum noise (environment) \cite{Paz-SilvaViola2014}.
In other words, our method can be applied to systems with both classical
noise and quantum noise present simultaneously.

\section{Ensemble average Infidelity and optimization method}

We first introduce our robust control method here, and then compare
it with the QSN method and the FTF method. We consider a total Hamiltonian
$\mathcal{H}(t)=\mathcal{H}_{I}(t)+\mathcal{H}_{N}(t)$, where $\mathcal{H}_{I}(t)$
is the ideal system Hamiltonian and $\mathcal{H}_{N}(t)$ is the noise
Hamiltonian. If a system is ideal, i.e., $\mathcal{H}_{N}(t)=0$,
then its propagator is $U_{I}(t)=\mathcal{T}_{+}\exp[-i\int_{0}^{t}\mathcal{H}_{I}(t')dt']$
(throughout this paper we set $\hbar=1$), where $\mathcal{T}_{+}$
is the time-ordering operator. However, in reality there may be many
sources of noise present in the system, so $\mathcal{H}_{N}(t)=\sum_{j}\beta_{j}(t)H_{N_{j}}(t)$,
where $\beta_{j}(t)$ is the strength of the $j$-th stochastic time-varying
noise and $H_{N_{j}}(t)$ is the corresponding system coupling operator
term. The propagator for a realistic system is then $U(t)=U_{I}(t)\cdot\mathcal{T}_{+}\exp[-i\int_{0}^{t}\tilde{\mathcal{H}}_{N}(t')dt']$.
Here $\tilde{\mathcal{H}}_{N}(t)=\Sigma_{j}\beta_{j}(t)R_{j}(t)$,
is the noise Hamiltonian in the interaction picture transformed by
$U_{I}(t)$ and $R_{j}(t)\equiv U_{I}^{\dagger}(t)H_{N_{j}}(t)U_{I}(t)$.
Suppose that $U_{T}$ is our target gate and gate operation time is
$t_{f}$. The gate infidelity (error) $\mathcal{I}$ for an $n$-qubit
gate can be defined as 
\begin{equation}
\mathcal{I}\equiv1-\dfrac{1}{4^{n}}\left|{\rm Tr}\left[U_{T}^{\dagger}U(t_{f})\right]\right|^{2},\label{eq:infidelity definition}
\end{equation}
where ${\rm Tr}$ denotes a trace over the $n$-qubit system state
space. If noise strength is not too strong, we can expand the propagator
$U(t_{f})$ in terms of $\tilde{\mathcal{H}}_{N}(t)$ by Dyson series
\cite{Dyson1949} into the form $U(t_{f})=U_{I}(t_{f})\cdot[I+\Psi_{1}+\Psi_{2}+\cdots]$,
where the first two terms of $\Psi_{j}$ are $\Psi_{1}=-i\int_{0}^{t_{f}}\tilde{\mathcal{H}}_{N}(t')dt'$,
and $\Psi_{2}=-\int_{0}^{t_{f}}dt_{1}\int_{0}^{t_{1}}dt_{2}\tilde{\mathcal{H}}_{N}(t_{1})\tilde{\mathcal{H}}_{N}(t_{2})$.
Substituting the expanded $U(t_{f})$ into $\mathcal{I}$ in Eq.~(\ref{eq:infidelity definition}),
the expanded infidelity $\mathcal{I}$ (see Appendix \ref{sec:appendix_a})
takes the form 
\begin{align}
\mathcal{I} & =J_{1}+J_{2}+\epsilon+\mathcal{O}(\tilde{\mathcal{H}}_{N}^{m},m\geq3),\label{eq:expanded infidelity}\\
J_{1} & \equiv1-\dfrac{1}{4^{n}}\left|{\rm Tr}\left[U_{T}^{\dagger}U_{I}(t_{f})\right]\right|^{2},\label{eq:J1}\\
J_{2} & \equiv-\dfrac{1}{2^{n-1}}{\rm Re}\left[{\rm Tr}\left(\Psi_{2}\right)\right]-\dfrac{1}{4^{n}}\left|{\rm Tr}\left(\Psi_{1}\right)\right|{}^{2}.\label{eq:J2}
\end{align}
Here $J_{1}$ is the definition of gate infidelity for the ideal system,
$J_{2}$ is the lowest-order contribution of the noise to the gate
infidelity, $\epsilon$ (detailed form shown in Appendix \ref{sec:appendix_a})
denotes an extra contribution that is correlated to $J_{1}$ and the
Dyson expansion terms $\Psi_{j}$, and $\mathcal{O}(\tilde{\mathcal{H}}_{N}^{m},m\geq3)$
represents other higher-order terms excluding $\epsilon$.
If noise strength is not too strong such that $|\Psi_{j+1}|\ll|\Psi_{j}|$,
the extra contribution $\epsilon$ will become negligible when $J_{1}$
is getting small (see discussion in Appendix \ref{sec:appendix_a}).
The symbol Re in Eq.~(\ref{eq:J2}) denotes taking the real part
of the quantity it acts on. Because noise $\beta_{j}(t)$ is stochastic,
we denote the ensemble average of the infidelity over the different
noise realizations as 
\begin{equation}
\left\langle \mathcal{I}\right\rangle =J_{1}+\left\langle J_{2}\right\rangle +\left\langle \epsilon\right\rangle +\left\langle \mathcal{O}(\tilde{\mathcal{H}}_{N}^{m},m\geq3)\right\rangle .\label{eq:ensemble average infidelity}
\end{equation}
Here 
\begin{eqnarray}
\left\langle J_{2}\right\rangle  & = & \underset{j,k}{\sum}\int_{0}^{t_{f}}dt_{1}\int_{0}^{t_{1}}dt_{2}C_{jk}(t_{1},t_{2})\dfrac{{\rm Tr}\left[R_{j}(t_{1})R_{k}(t_{2})\right]}{2^{n-1}}\nonumber \\
 &  & \hspace{-0.5cm}-\underset{j,k}{\sum}\int_{0}^{t_{f}}dt_{1}\int_{0}^{t_{f}}dt_{2}C_{jk}(t_{1},t_{2})\dfrac{{\rm Tr}\left[R_{j}(t_{1})\right]{\rm Tr}\left[R_{k}(t_{2})\right]}{4^{n}},\label{eq:ensemble average j2}
\end{eqnarray}
where $C_{jk}(t_{1},t_{2})=\left\langle \beta_{j}(t_{1})\beta_{k}(t_{2})\right\rangle $
is the CF for noise $\beta_{j}(t_{1})$ and $\beta_{k}(t_{2})$. The
first-order noise term proportional to ${\rm Re}[{\rm Tr}(\Psi_{1})]$
vanishes due to the fact that ${\rm Tr}(\Psi_{1})$ is purely imaginary
rather than the assumption of $\left\langle \beta_{j}(t)\right\rangle =0$
%There%are no $\left\langle \beta_{j}(t)\right\rangle $ terms in%Eq. (\ref{eq:ensemble  average j2})%for ${\rm Re}\left[{\rm Tr}\left(\Psi_{1}\right)\right]=0$, not because%of the assumption of $\left\langle \beta_{j}(t)\right\rangle =0$(see
(see Appendix \ref{sec:appendix_a}). If different sources of noise are
independent, $C_{jk}(t_{1},t_{2})=0$ for $j\neq k$, and if noise
Hamiltonian $\mathcal{H}_{N}(t)$ is traceless, the second term in
Eq.~(\ref{eq:ensemble average j2}) vanishes. The ideal Hamiltonian
$\mathcal{H}_{I}(t)$ is a function of the control field $\Omega(t),$
that is $\mathcal{H}_{I}(t)=\mathcal{H}_{I}(\Omega(t))$, and the
control field $\Omega(t)$ is chosen to be a function of a set of
control parameters $[a_{1},a_{2},\cdots]$. Then $U_{I}(t)$ and each
term of the ensemble average infidelity $\left\langle \mathcal{I}\right\rangle $
in Eq.~(\ref{eq:ensemble average infidelity}) are also a function
of the control parameter set $[a_{1},a_{2},\cdots]$. Our goal is
to search the optimal parameter set $[a_{1},a_{2},\cdots]$ that minimizes
the ensemble average infidelity $\left\langle \mathcal{I}\right\rangle $.
If the noise strength or fluctuation is not large, then the dominant
noise contribution to $\left\langle \mathcal{I}\right\rangle $ is
from $\left\langle J_{2}\right\rangle $ as the higher order terms
$\langle\mathcal{O}(\tilde{\mathcal{H}}_{N}^{m},m\geq3)\rangle$ can
be neglected (see Appendix~\ref{sec:appendix_b}). %We provide an estimation for how small %method to determine how small%noise strength is required for neglecting $\langle\mathcal{O}(\tilde{\mathcal{H}}_{N}^{m},m\geq3)\rangle$%in Appendix~\ref{sec:appendix_b}. Besides,
$J_{1}$ can generally be made sufficiently small so that the extra
term $\left\langle \epsilon\right\rangle $ in $\left\langle \mathcal{I}\right\rangle $
of Eq.~(\ref{eq:ensemble average infidelity}) can be safely ignored.
So we concentrate on the minimization of $\left\langle \mathcal{I}\right\rangle \cong J_{1}+\left\langle J_{2}\right\rangle $
for obtaining the optimal control parameter set. We will, however,
use the full-order ensemble average infidelity $\left\langle \mathcal{I}\right\rangle $
(described later) to exam the performance of the optimal control parameter
set found this way.

We use two-step optimization to achieve this goal. The first step
is called the $J_{1}$ optimization in which $J_{1}$ is the cost
function. The gate infidelities $J_{1}$ in an ideal unitary system
with gate-operation-controllability and a sufficient number of control
parameters can be made as low as one wishes, limited only by the machine
precision of the computation. So using an ensemble of random control
parameter sets as initial guesses, we obtain after the $J_{1}$ optimization
an ensemble of optimized control parameters sets all with very low
values of $J_{1}$. The second step is called the $J_{1}+\left\langle J_{2}\right\rangle $
optimization. We take $J_{1}+\left\langle J_{2}\right\rangle $ as
a cost function and randomly choose some optimized control parameter
sets in the first optimization step as initial guesses to run the
optimal control algorithm. After the $J_{1}+\left\langle J_{2}\right\rangle $
optimization, we obtain an ensemble of control parameter sets with
low values of $J_{1}+\left\langle J_{2}\right\rangle $, and then
choose the lowest one as the optimal control parameter set. The purpose
of using the two-step optimization is to improve optimization efficiency.
If we run $J_{1}+\left\langle J_{2}\right\rangle $ optimization directly
from an ensemble of random control parameter sets, we need more optimization
iterations to achieve the goal, and the success rate is relatively
low compared with the two-step optimization. Besides, the $J_{1}+\left\langle J_{2}\right\rangle $
optimization enables us to know separately the optimized values of
$J_{1}$ and $\left\langle J_{2}\right\rangle $. When $\left\langle J_{2}\right\rangle $
can be minimized to a very small value as in the case of static or
low-frequency noise, one has to use a small time step for simulation
to make $J_{1}$ smaller than $\left\langle J_{2}\right\rangle $.
However, for high-frequency noise, $\left\langle J_{2}\right\rangle $
is hard to be minimized to a very small value, and one can instead
choose a suitable larger time step to make $J_{1}$ just one or two
orders of magnitude smaller than $\left\langle J_{2}\right\rangle $,
saving substantially the optimization time especially for multi-qubits
and multi-sources of noise. We use the gradient-free and model-free
Nelder-Mead (NM) algorithm \cite{NelderMead1965} in both the $J_{1}$
and $J_{1}+\left\langle J_{2}\right\rangle $ optimization steps.
However, the NM algorithm may be stuck in local traps in the $J_{1}+\left\langle J_{2}\right\rangle $
parameter space topography. To overcome this problem, we use repeating-NM
algorithm in the $J_{1}+\left\langle J_{2}\right\rangle $ optimization
step. The control parameter set from the first $J_{1}+\left\langle J_{2}\right\rangle $
optimization may lie in a local trap. Therefore, we add random fluctuations
to this control parameter set and try to pull it out of the trap.
Then we use this shifted control parameter set as an initial guess
to run the second $J_{1}+\left\langle J_{2}\right\rangle $ optimization.
We repeat the same procedure many times until the values of $J_{1}+\left\langle J_{2}\right\rangle $
can not be improved (reduced) anymore, and then output the corresponding
control parameter set. Our optimization method employing the gradient-free
and model-free NM algorithm is quite general, capable of dealing with
different forms or structures of the ideal system Hamiltonian $\mathcal{H}_{I}(t)$,
control field $\Omega(t)$, noise Hamiltonian $\mathcal{H}_{N}(t)$,
and noise CF $C_{jk}(t_{1},t_{2})$ for a few qubit systems.

The ensemble infidelity $\left\langle \mathcal{I}\right\rangle $
we use to show the robust performance of the gate as the noise strength
varies is calculated using the full evolution of the total system-noise
Hamiltonian and many realizations of the noise without any other approximation.
By inputting the optimal control parameter set obtained by the optimization
strategy into the total system-noise Hamiltonian $\mathcal{H}(t)=\mathcal{H}_{I}(t)+\mathcal{H}_{N}(t)$
to obtain numerically the full propagator for a single noise realization,
we can calculate the gate infidelity $\mathcal{I}$ using Eq.~(\ref{eq:infidelity definition})
for the noise realization. The procedure is repeated for many different
noise realizations. Then we take an ensemble average of the infidelities
over the different noise realizations to obtain $\left\langle \mathcal{I}\right\rangle $.

In principle, we could deal with any given form of the noise correlation
function (or equivalently the noise PSD) to inset into Eq.~(\ref{eq:ensemble average j2})
for the $J_{1}+\left\langle J_{2}\right\rangle $ optimization. But
as a particular example, we choose the Ornstein-Uhlenbeck (OU) process
$\beta_{OU}(t)$ to simulate stochastic time-varying noise \cite{FinchYt2004}.
Studying the influence of and developing robust strategies against
time-dependent noise is an important subject of research in quantum
control problems both theoretically and experimentally \cite{SoareBallHayesEtAl2014,BallBiercuk2015,GreenSastrawanUysEtAl2013,Oliver2014,KabytayevGreenKhodjastehEtAl2014}.
If the initial noise $\beta_{OU}(t=0)$ is a normal distribution with
zero mean and with standard deviation $\sigma_{OU}$, then the noise
CF of the OU process $\beta_{OU}(t)$ is 
\begin{equation}
C_{OU}(t_{1},t_{2})=\sigma_{OU}^{2}\exp\left(-\gamma_{OU}\left|t_{1}-t_{2}\right|\right)\label{eq:ou correlation function}
\end{equation}
with the noise correlation time $\tau\sim(1/\gamma_{OU})$, and the
corresponding noise PSD is Lorentzian 
\begin{equation}
S_{OU}(\omega)=\frac{2\sigma_{OU}^{2}\gamma_{OU}}{(\gamma_{OU}^{2}+\omega^{2})}.\label{eq:OU_PSD}
\end{equation}
Lorentzian PSDs of spin noise resulting in a fluctuating magnetic
field at the location of the qubits in InGaAs semiconductor quantum
dots have been measured experimentally \cite{KuhlmannHouelLudwigEtAl2013,LiSinitsynSmithEtAl2012}.
Generally, small $\gamma_{OU}$ corresponds to low-frequency or quasi-static
noise; large $\gamma_{OU}$ corresponds to high-frequency noise. The
noise $\beta_{OU}(t)$ can be simulated through the formula $\beta_{OU}(t+dt)=\left(1-\gamma_{OU}dt\right)\beta_{OU}(t)+\sigma_{OU}\sqrt{2\gamma_{OU}}dW(t)$,
where $W(t)$ is a Wiener process \cite{FinchYt2004}. Figures \ref{Fig1}(c),
(d), and (e) show the different realizations of the noise $\beta_{OU}(t)$
with $\sigma_{OU}=10^{-3}$ for different values of $\gamma_{OU}/\omega_{0}=10^{-7}$,
$10^{-3}$, and $10^{-1}$, respectively, where $\omega_{0}$ is the
typical system frequency. We note here that the particular choice
of the OU noise should by no means diminish the value of our work
and the power of our method. Any given or experimentally measured
well-behaved noise PSD or noise correlation function can be dealt
with. We will demonstrate later that our method can also work effectively
for another form of noise PSD different from that of the OU noise
when we compare the performance of our method with the FTF method.
The reason to use the OU noise in the system-noise Hamiltonian here
is that it is relatively easy to simulate its stochastic noise realizations
in the time domain. Therefore, we can calculate the full-order ensemble
average infidelity $\left\langle \mathcal{I}\right\rangle $ to show
that our $J_{1}+\left\langle J_{2}\right\rangle $ optimization that
minimizes the second-order noise contribution to the average infidelity
$\left\langle \mathcal{I}\right\rangle $ can indeed work rather well
for not too strong a noise fluctuation.

\section{Results and demonstrations}

\subsection{Comparison with quasi-static-noise method}

\subsubsection{Single-qubit gates}

\label{sub:one_qubit}

\begin{figure}
\onlinecap\includegraphics[scale=0.55]{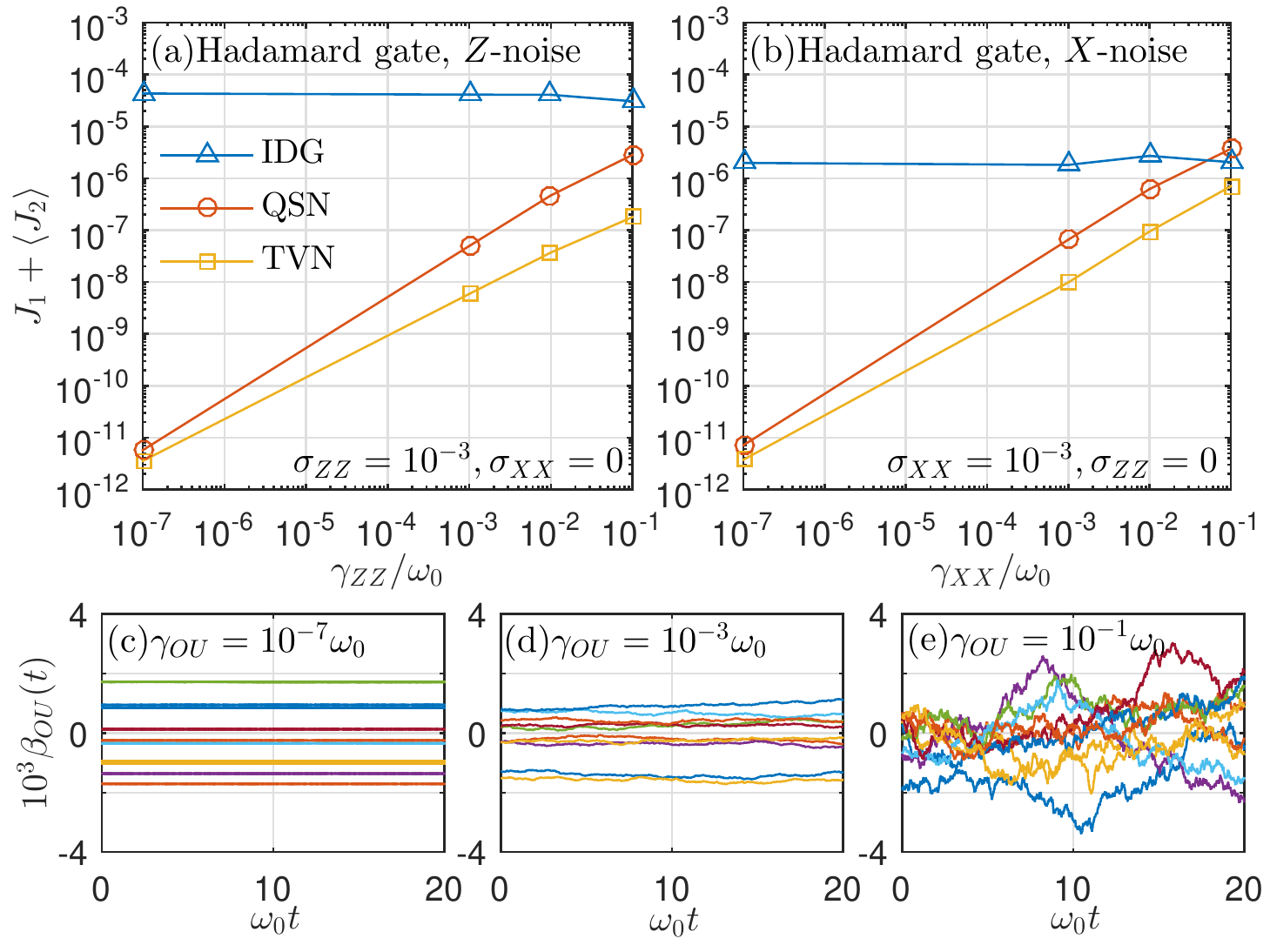} \protect\protect\protect\caption{$J_{1}+\left\langle J_{2}\right\rangle $ versus (a) $\gamma_{ZZ}$
for $Z$-noise ($\sigma_{ZZ}=10^{-3}$, $\sigma_{XX}=0$) and (b)
$\gamma_{XX}$ for $X$-noise ($\sigma_{XX}=10^{-3}$, $\sigma_{ZZ}=0$).
The $J_{1}+\left\langle J_{2}\right\rangle $ values are obtained
using the optimal control parameter sets of Hadamard gate from the
IDG strategy (in blue triangles), QSN strategy (in orange circles),
and TVN strategy (in yellow squares). (c), (d), and (e) are 10 realizations
of OU noise $\beta_{OU}(t)$ for $\gamma_{OU}/\omega_{0}=10^{-7}$,
$10^{-3}$, $10^{-1}$ , and $\sigma_{OU}=10^{-3}$.}

\label{Fig1} 
\end{figure}

\begin{figure}
\onlinecap\includegraphics[scale=0.60]{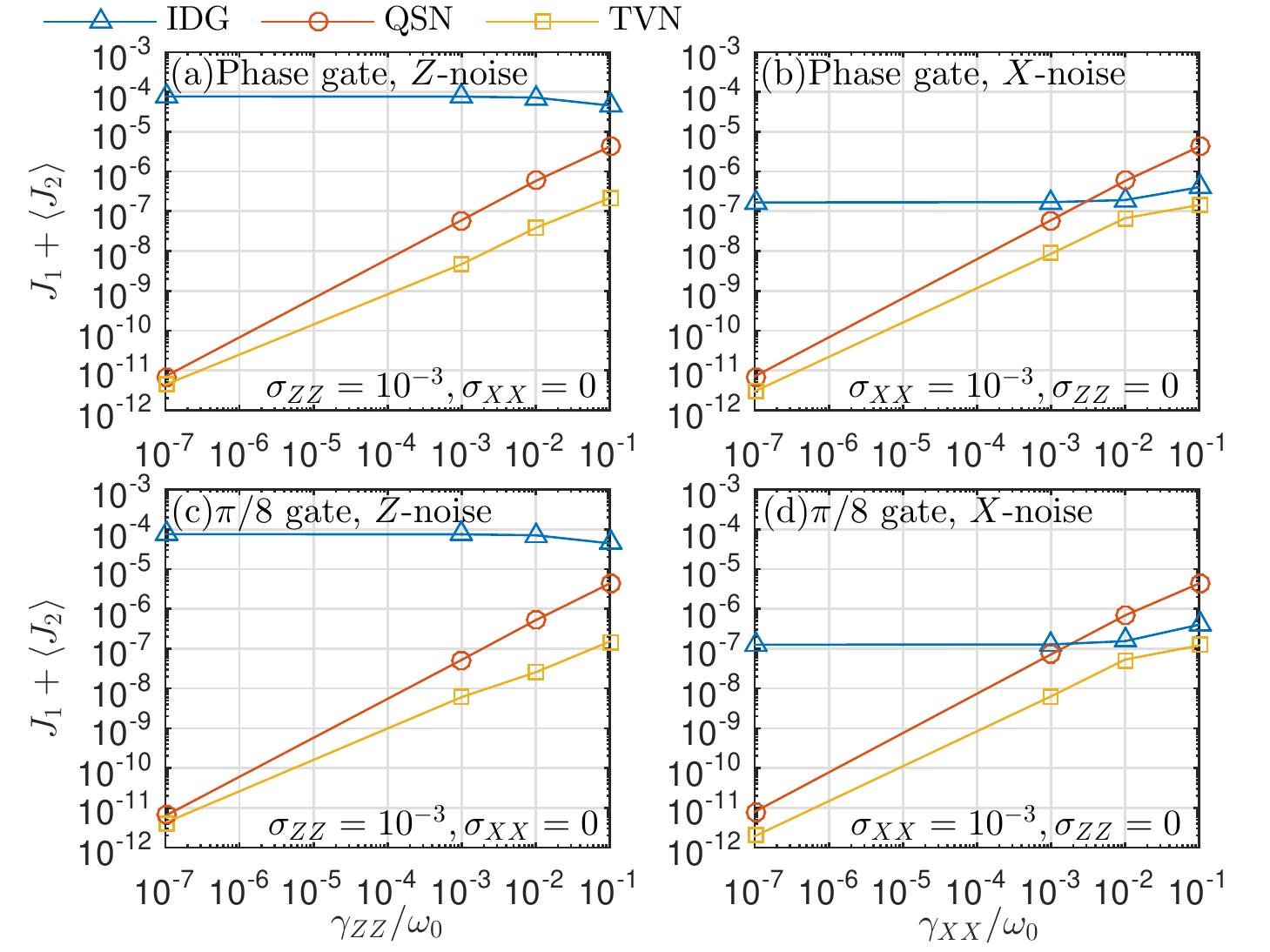}\protect \protect\protect\caption{ $J_{1}+\left\langle J_{2}\right\rangle $ values versus $\gamma_{ZZ}$
for $Z$-noise ($\sigma_{ZZ}=10^{-3}$, $\sigma_{XX}=0$) and versus
$\gamma_{XX}$ for $X$-noise ($\sigma_{XX}=10^{-3}$, $\sigma_{ZZ}=0$)
obtained from the IDG strategy (in blue triangles), QSN strategy (in
orange circles), and TVN strategy (in yellow squares) for the phase
gate shown in (a) and (b) respectively, and for the $\pi/8$ gate
in (c) and (d), respectively. }

\label{Fig_add} 
\end{figure}

We demonstrate as an example the implementation of single-qubit gates
in the presence of time-varying noise using our method. The ideal
system Hamiltonian for the qubit is 
\begin{equation}
\mathcal{H}_{I}(t)=\omega_{0}\frac{Z}{2}+\Omega_{X}(t)\frac{X}{2},\label{eq:1qubit_H}
\end{equation}
where $X$ and $Z$ stand for the Pauli matrices, $\omega_{0}$ is
the qubit transition frequency, and $\Omega_{X}(t)$ is the control
field in the $X$ term. The noise Hamiltonian is written as 
\begin{equation}
\mathcal{H}_{N}(t)=\beta_{Z}(t)\omega_{0}\frac{Z}{2}+\beta_{X}(t)\Omega_{X}(t)\frac{X}{2}.\label{eq:Noise_H}
\end{equation}
We call $\beta_{Z}(t)$ the $Z$-noise and $\beta_{X}(t)$ the $X$-noise,
and assume that they are independent OU noises with CF's $C_{ZZ}(t_{1},t_{2})=\sigma_{ZZ}^{2}\exp\left(-\gamma_{ZZ}\left|t_{1}-t_{2}\right|\right)$
and $C_{XX}(t_{1},t_{2})=\sigma_{XX}^{2}\exp\left(-\gamma_{XX}\left|t_{1}-t_{2}\right|\right)$
as the form of Eq.~(\ref{eq:ou correlation function}). We choose
the control pulse as a composite sine pulse expressed as 
\begin{equation}
\Omega_{X}(t)=\sum_{k=1}^{k_{{\rm max}}}a_{k}\sin\left(m_{k}\pi\frac{t}{t_{f}}\right),\label{eq:composite_sine_pulse}
\end{equation}
where the set of the strengths of the single sine pulses is the control
parameter set $[a_{k}]=[a_{1},a_{2},\cdots,a_{k_{{\rm max}}}]$ and
$\{m_{k}\}$ is a set of integers, chosen depending on the nature
of the system Hamiltonians and the target gates as well as the properties
of the noise models. For each control pulse, we choose the number of control parameters $k_{{\rm max}}$
to range from $8$ to $20$ in our calculations.

We define below three optimization strategies, namely, the ideal-gate
(IDG) strategy, quasi-static-noise (QSN) strategy, and time-varying-noise
(TVN) strategy. The IDG strategy is to perform the first-step optimization
($J_{1}$ optimization) only and to show the performance of an ideal
gate pulse in the presence of noise. The TVN strategy is our proposed
method described earlier above, in which the actual $\gamma_{ZZ}$
and $\gamma_{XX}$ values are used in the noise CF's of the cost function
$\left\langle J_{2}\right\rangle $ for the second-step optimization.
The QSN strategy uses the same optimization procedure as the TVN strategy,
but with $\gamma_{ZZ}=\gamma_{XX}=0$ for the noise CF's in the cost
function $\left\langle J_{2}\right\rangle $. Thus it is regarded
to represent the QSN methods. We choose the gate operation time $t_{f}=20/\omega_{0}$.
After the optimizations of Hadamard gate, we plot the corresponding
$J_{1}+\left\langle J_{2}\right\rangle $ values obtained from these
three strategies versus $\gamma_{ZZ}$ in Fig.~\ref{Fig1}(a) for
the $Z$-noise and versus $\gamma_{XX}$ in Fig.~\ref{Fig1}(b) for
the $X$-noise. For low-frequency (quasi-static) noise ($\gamma_{ZZ}=\gamma_{XX}=10^{-7}\omega_{0}$),
the performance of the TVN strategy and the QSN strategy are about
the same but they are several orders of magnitude better in infidelity
$J_{1}+\left\langle J_{2}\right\rangle $ value than the IDG strategy
which does not take the noise into account at all. As the noise goes
from the low frequency to high frequency ($\gamma_{ZZ}=\gamma_{XX}=10^{-1}\omega_{0}$),
the TVN strategy taking account of the time-varying noise information
in the cost function gets better and better (from a factor-level to
an order-of-magnitude-level) improvement in $J_{1}+\left\langle J_{2}\right\rangle $
values than the QSN strategy in which noise is assumed to be quasi-static.
In addition to the Hadamard gate, we perform calculations for other
quantum gates, namely the phase gate, $\pi/8$ gate and controlled-NOT
(CNOT) gate, in the fault-tolerant universal set in terms of which
any unitary operation can be expressed to arbitrary accuracy. The
$J_{1}+\left\langle J_{2}\right\rangle $ values versus $\gamma_{ZZ}$
and versus $\gamma_{XX}$ obtained from the three strategies are shown
in Figs.~\ref{Fig_add}(a) and (b), respectively, for the phase gate
and in Figs.~\ref{Fig_add}(c) and (d), respectively, for the $\pi/8$
gate. Their performances are similar to those in Fig.~\ref{Fig1}(a)
and (b) of the Hadamard gate. The optimization results for the two-qubit
CNOT gate are presented in Sec.~\ref{sec:two-qubit}.

\begin{figure}
\onlinecap \includegraphics[scale=0.55]{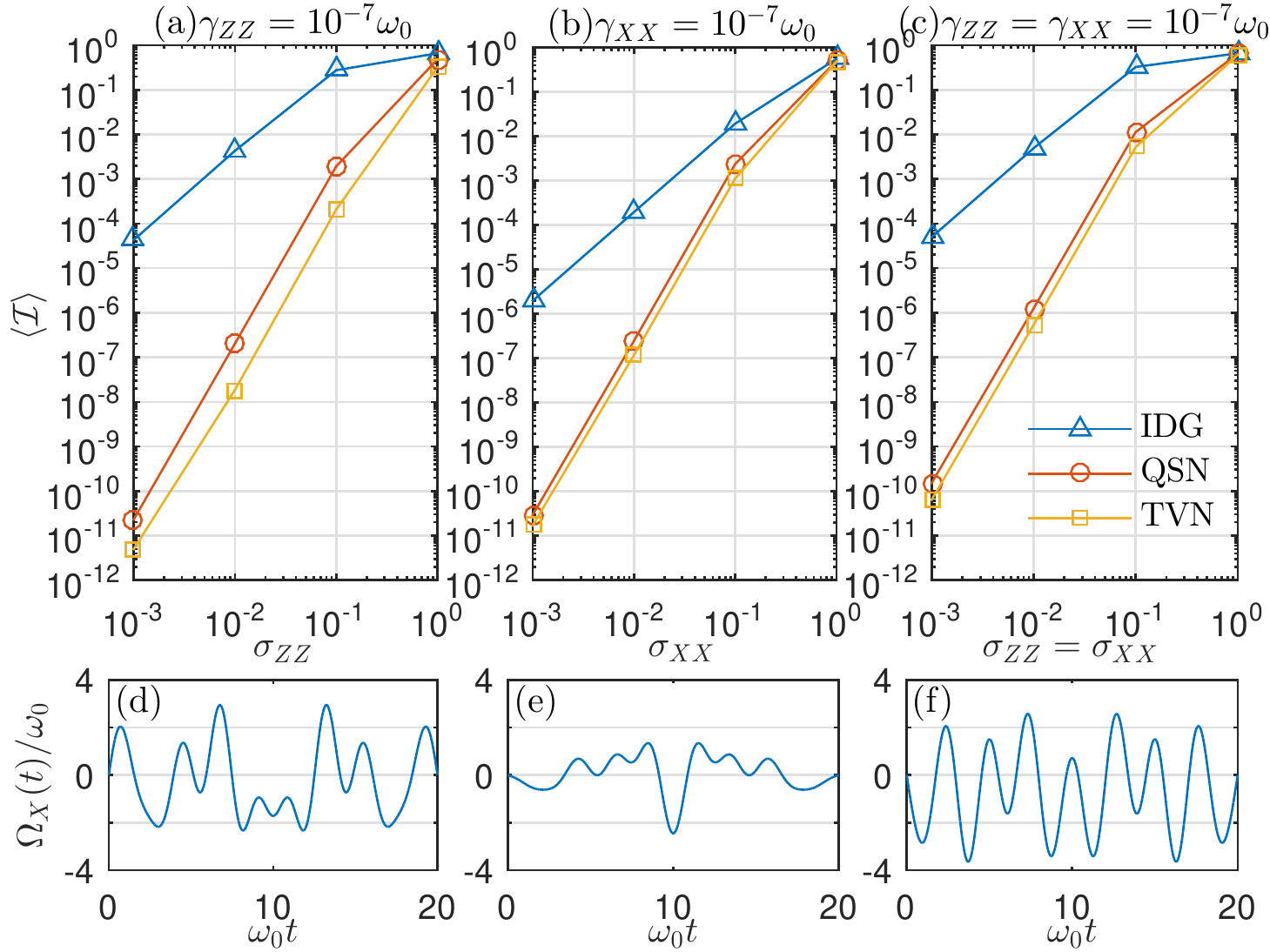}\protect\protect\protect\caption{Robust performance of the Hadamard gate of the IDG strategy (in blue
triangles), QSN strategy (in orange circles), and TVN strategy (in
yellow squares) for low-frequency ($\gamma_{ZZ}=\gamma_{XX}=10^{-7}\omega_{0}$)
(a) $Z$-noise, (b) $X$-noise, and (c) $Z$-$\&$-$X$-noise. The
corresponding optimal control pulses of the TVN strategy for $Z$-noise,
$X$-noise, and $Z$-$\&$-$X$-noise are shown in (d), (e), and (f),
respectively. The number of control parameters $k_{\rm{max}}$=10 for $\Omega_{X}(t)$ in (d), (e), and (f).}

\label{Fig2} 
\end{figure}

\begin{figure}
\onlinecap \includegraphics[scale=0.55]{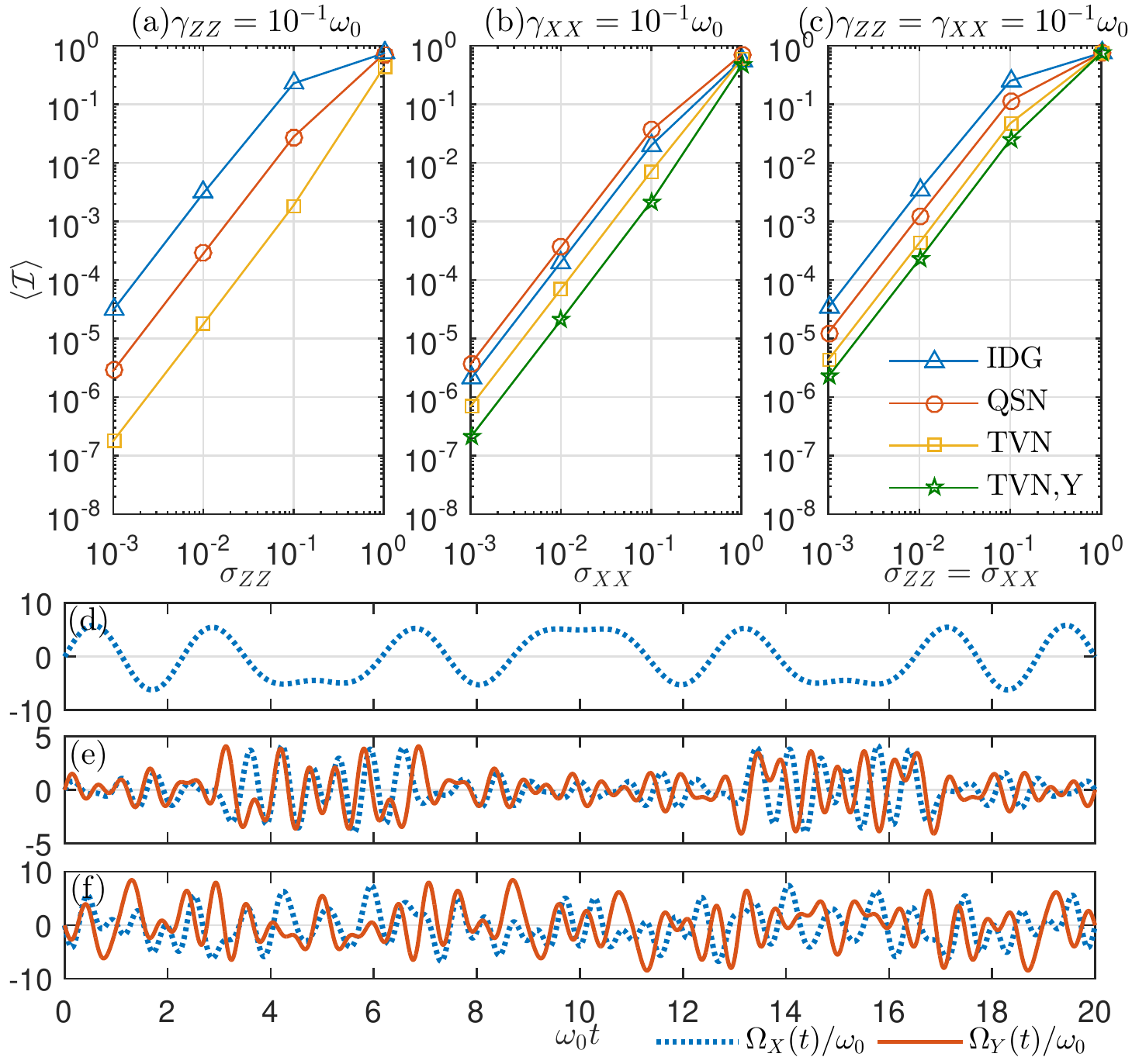}\protect\protect\protect\caption{Robust performance of the Hadamard gate of the IDG strategy (in blue
triangles), QSN strategy (in orange circles), and TVN strategy (in
yellow squares) for high-frequency ($\gamma_{ZZ}=\gamma_{XX}=10^{-1}\omega_{0}$)
(a) $Z$-noise, (b) $X$-noise, and (c) $Z$-$\&$-$X$-noise. For
TVN strategy with an additional $Y$ control (in green pentagrams)
in (b), $\gamma_{YY}=\gamma_{XX}=10^{-1}\omega_{0}$ and $\sigma_{YY}=\sigma_{XX}$,
and in (c), $\gamma_{YY}=\gamma_{ZZ}=\gamma_{XX}=10^{-1}\omega_{0}$
and $\sigma_{YY}=\sigma_{XX}=\sigma_{ZZ}$. The optimal control pulses
of the TVN strategy for the $Z$-noise is shown in (d) and of TVN
strategy with an additional $Y$ control for the $X$-noise and the
$Z$-$\&$-$X$-noise with accompanying $Y$-noise are shown in (e)
and (f), respectively. The numbers of control parameters $k_{\rm{max}}$=10 for $\Omega_{X}(t)$ in (d), and $k_{\rm{max}}$=20 for $\Omega_{X}(t)$ and $\Omega_{Y}(t)$ in (e) and (f).}

\label{Fig3} 
\end{figure}

Next, we take the optimal control parameter sets of Hadamard gate
from these three strategies to show their robust performance against
$Z$-noise, $X$-noise, and $Z$-\&-$X$-noise at a low frequency
($\gamma_{ZZ}=\gamma_{XX}=10^{-7}\omega_{0}$) in Figs.~\ref{Fig2}(a),
(b), and (c) and at a high frequency ($\gamma_{ZZ}=\gamma_{XX}=10^{-1}\omega_{0}$)
in Figs.~\ref{Fig3}(a), (b), and (c). For low-frequency noises and
for small noise strength ($\sigma_{XX}<10^{-1}$, $\sigma_{ZZ}<10^{-1}$),
one can see from Fig.~\ref{Fig2} that the full-order ensemble average
infidelity $\left\langle \mathcal{I}\right\rangle $ scales for the
IDG strategy as the second power of the noise standard deviation ($\sigma_{ZZ}$,
$\sigma_{XX}$), but scales for the TVN and QSN strategies as the
fourth power. This implies that $\left\langle \mathcal{I}\right\rangle \cong\left\langle J_{2}\right\rangle $
for the IDG strategy, but the TVN and QSN strategies can nullify the
contribution from $\left\langle J_{2}\right\rangle $ for the low-frequency
(quasi-static) noise and the dominant contribution in $\left\langle \mathcal{I}\right\rangle $
comes from the next higher-order term, i.e., $\langle\mathcal{I}\rangle\cong\langle\mathcal{O}(\tilde{\mathcal{H}}_{N}^{4})\rangle$.
In this case, our method, the TVN strategy, still performs slightly
better than the QSN strategy. For gate error (infidelity) less than
the error threshold of $10^{-2}$ of surface codes \cite{FowlerMariantoniMartinisEtAl2012}
required for fault-tolerant quantum computation (FTQC), the Hadamard
gate of TVN strategy can be robust to $\sigma_{ZZ}\sim30\%$ for low-frequency
$Z$-noise (i.e., against noise fluctuation with standard deviation
up to about $30\%$ of $\omega_{0}/2$), robust to $\sigma_{XX}\sim20\%$
for the $X$-noise {[}i.e., against noise fluctuation with standard
deviation up to about $20\%$ of $\Omega_{X}(t)/2${]}, and robust
to $\sigma_{ZZ}=\sigma_{XX}\sim10\%$ for $Z$-$\&$-$X$-noise as
shown in Figs.~\ref{Fig2}(a), (b), and (c), respectively. The corresponding
optimal control pulses of the TVN strategy are shown in Figs.~\ref{Fig2}(d),
(e), and (f), respectively.

For high-frequency noise shown in Fig.~\ref{Fig3}, the full-order
ensemble average infidelity $\left\langle \mathcal{I}\right\rangle $
scales as the second power of noise standard deviation ($\sigma_{ZZ}$,
$\sigma_{XX}$) for all three strategies and noises. This indicates
that for high frequency noise $\left\langle J_{2}\right\rangle $
is not nullified completely, and is only minimized. Even in this case,
the TVN strategy still has over two orders of magnitude improvement
in $\left\langle \mathcal{I}\right\rangle $ compared with the IDG
strategy, and over one order of magnitude improvement compared with
the QSN strategy for the $Z$-noise at small noise strengths as shown
in Fig.~\ref{Fig3}(a). For $\left\langle \mathcal{I}\right\rangle \lesssim10^{-2}$
less than the FTQC error threshold of the surface codes, the Hadamard
gate implemented by our optimal control pulse shown in Fig.~\ref{Fig3}(d)
can be robust to $\sigma_{ZZ}\sim20\%$ for the $Z$-noise. On the
other hand, for the high-frequency $X$-noise, $\left\langle \mathcal{I}\right\rangle $
obtained by the QSN strategy has even slightly higher values than
those by the IDG strategy. The improvement in $\left\langle \mathcal{I}\right\rangle $
by the TVN strategy over the other two strategies is less than one
order of magnitude. To improve the gate performance, we increase the
degrees of freedom for optimization by adding a control term $\Omega_{Y}(t)Y/2$
and its accompanying $Y$-noise term $\beta_{Y}(t)\Omega_{Y}(t)Y/2$
in the Hamiltonian. We choose, for simplicity, $\gamma_{YY}=\gamma_{XX}$
and $\sigma_{YY}=\sigma_{XX}$, and use the same optimal procedure
as the TVN strategy. The improvement in $\left\langle \mathcal{I}\right\rangle $
of the TVN strategy with an additional $Y$ control as compared with
the TVN strategy is over a half order of magnitude. As a result, the
Hadamard gate with the optimal control pulses of the TVN strategy
with an additional $Y$ control shown in Fig.~\ref{Fig3}(e) can
be robust to $\sigma_{XX}=\sigma_{YY}\sim20\%$ for $\left\langle \mathcal{I}\right\rangle \lesssim10^{-2}$.
Note that the optimization algorithm seems to find control pulses
with stronger strengths to suppress the $Z$-noise, but searches weaker
control pulses to minimize the $X$-noise cost function since the
system coupling operator term of the $X$-noise is proportional to
the control field $\Omega_{X}(t)$ in our noise model. So for the
case with the $Z$-noise and $X$-noise simultaneously present, there
is a trade-off in the control pulse strength for the cost function
optimization between the $Z$-noise and the $X$-noise. Consequently,
the ensemble infidelity of the $Z$-$\&$-$X$ noise does not reach
a low value as the case with only the $Z$-noise or the $X$-noise.
Thus one can see from Fig.~\ref{Fig3}(c) that the improvement in
$\left\langle \mathcal{I}\right\rangle $ of the TVN strategy over
the IDG strategy is just near one order of magnitude, and only a half
order as compared with the QSN strategy. Similar trade-off also takes
place for the TVN strategy with additional $Y$ control although it
performs slightly better than the TVN strategy with only the $\Omega_{X}(t)$
control field. Nevertheless, the Hadamard gate implemented with the
optimal pulse obtained by the TVN strategy with additional $Y$ control
shown in Fig.\ \ref{Fig3}(f) can be still robust to $\sigma_{ZZ}=\sigma_{XX}=\sigma_{YY}\sim6\%$
for $\left\langle \mathcal{I}\right\rangle \lesssim10^{-2}$.

\subsubsection{Two-qubit gates}

\label{sec:two-qubit}

\begin{figure}
\onlinecap \includegraphics[scale=0.60]{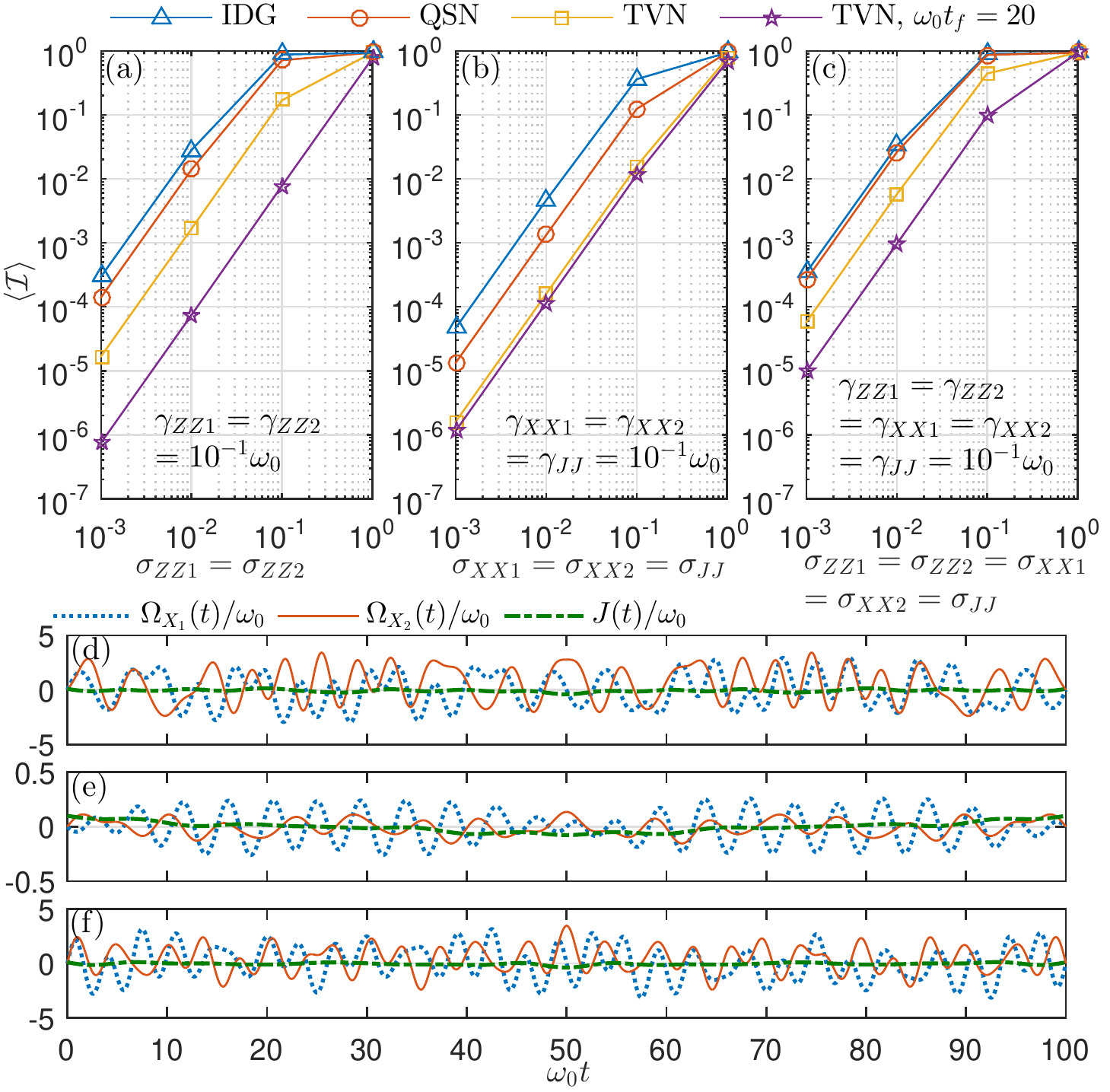}\protect\protect\protect\caption{Robust performance of CNOT gates of the IDG strategy ($\omega_{0}t_{f}=100$
in blue triangles), QSN strategy ($\omega_{0}t_{f}=100$ in orange
circles), TVN strategy ($\omega_{0}t_{f}=100$ in yellow squares and
$\omega_{0}t_{f}=20$ in purple pentagrams) for high-frequency ($\gamma_{ZZ1}=\gamma_{ZZ2}=\gamma_{XX1}=\gamma_{XX2}=\gamma_{JJ}=10^{-1}\omega_{0}$)
(a) $Z$-noise, (b) $X$-\&-$J$-noise, and (c) $Z$-\&-$X$-\&-$J$-noise.
The optimal control pulses $\Omega_{X_{1}}(t)$ in thick dotted blue line, $\Omega_{X_{2}}(t)$ in thin solid orange line, and $J(t)$ in thick dash-dotted green line of the TVN strategy ($\omega_{0}t_{f}=100$)
for the $Z$-noise, $X$-\&-$J$-noise, and $Z$-\&-$X$-\&-$J$-noise
are shown in (d), (e), and (f), respectively. The numbers of control parameters $k_{\rm{max}}$=16, 16, and 8 for $\Omega_{X_{1}}(t)$, $\Omega_{X_{2}}(t)$, and $J(t)$ respectively in (d) and (f); $k_{\rm{max}}$=12, 12, and 6 for $\Omega_{X_{1}}(t)$, $\Omega_{X_{2}}(t)$, and $J(t)$ respectively in (e).}

\label{sFig1} 
\end{figure}

Next, we demonstrate that our method can find control pulses for high-fidelity
two-qubit CNOT gate operations in the presence of multi-sources of
high-frequency noise. The two-qubit Hamiltonian is chosen ass 
\begin{equation}
\mathcal{H}_{I}(t)=\omega_{0}\frac{Z_{1}}{2}+\Omega_{X_{1}}(t)\frac{X_{1}}{2}+\omega_{0}\frac{Z_{2}}{2}+\Omega_{X_{2}}(t)\frac{X_{2}}{2}+J(t)\frac{Z_{1}Z_{2}}{2},\label{eq:2qubit_H}
\end{equation}
where $Z_{j}$ and $X_{j}$ denote the Pauli's matrix operators for
qubit $j$, $\Omega_{X_{j}}(t)$ is the control field applied to qubit
$j$ and $J(t)$ is the two-qubit coupling strength. We assume OU
noise can be present in each of the five terms, and $\sigma_{ZZ1}$,
$\sigma_{ZZ2}$, $\sigma_{XX1}$, $\sigma_{XX2}$, and $\sigma_{JJ}$
are respectively the corresponding standard deviation $\sigma_{OU}$,
and $\gamma_{ZZ1}$, $\gamma_{ZZ2}$, $\gamma_{XX1}$, $\gamma_{XX2}$,
and $\gamma_{JJ}$ are respectively the corresponding $\gamma_{OU}$.
We choose the control fields $\Omega_{X_{1}}(t)$ and $\Omega_{X_{2}}(t)$
as composite sine pulses, and the two-qubit control $J(t)$ as a composite
sine pulse with a constant shift.

The robust performance of the CNOT gate using the three strategies
for high frequency ($\gamma_{ZZ1}=\gamma_{ZZ2}=\gamma_{XX1}=\gamma_{XX2}=\gamma_{JJ}=10^{-1}\omega_{0})$
$Z$-noise, $X$-\&-$J$-noise, and $Z$-\&-$X$-\&-$J$-noise are
shown in Figs.~\ref{sFig1}(a), (b), and (c), respectively. The corresponding
optimal control pulses of the TVN strategy for operation time $t_{f}=100/\omega_{0}$
are shown in Figs.~\ref{sFig1}(d), (e), and (f), respectively. For
$\omega_{0}t_{f}=100$, our method (the TVN strategy) for the case
of the $Z$-noise and the case of the $X$-\&-$J$-noise has one-order
of magnitude improvement in $\left\langle \mathcal{I}\right\rangle $
values as compared with the QSN strategy for small noise strength,
but only half-order improvement for the case of the $Z$-\&-$X$-\&-$J$-noise.
This is because for the case of the $Z$-\&-$X$-\&-$J$-noise, there
is a trade-off in the control pulse strength for the cost function
optimization between the $Z$-noise and the $X$-\&-$J$-noise, similar
to that in the single-qubit case. The robust performance can be improved
by reducing gate operation time $t_{f}$, for example, from $t_{f}=100/\omega_{0}$
to $t_{f}=20/\omega_{0}$, to decrease the duration of the influence
of the noises. This can be seen from the purple pentagrams in Figs.~\ref{sFig1}(a)
and (c). For the case of the $X$-\&-$J$-noise in Fig.~\ref{sFig1}(b),
only slight improvement is observed for the $t_{f}=20/\omega_{0}$
case because when the operation time decreases, it is hard to make
the strengths of the control fields $\Omega_{X}^{j}(t)$ and $J(t)$
all small as in the $t_{f}=100/\omega_{0}$ case. For high-frequency
noise and for FTQC error threshold $\left\langle \mathcal{I}\right\rangle \le10^{-2}$
of the surface codes, the CNOT gate with operation time $t_{f}=20/\omega_{0}$
can be robust to $\sigma_{ZZ1}=\sigma_{ZZ2}\sim10\%$ for the $Z$-noise,
robust to $\sigma_{XX1}=\sigma_{XX2}=\sigma_{JJ}\sim10\%$ for the
$X$-\&-$J$-noise, 
and robust to
$\sigma_{ZZ1}=\sigma_{ZZ2}=\sigma_{XX1}=\sigma_{XX2}=\sigma_{JJ}\sim3\%$ 
for the $Z$-\&-$X$-\&-$J$-noise by our method.

We describe briefly about the computational resources and computation
time in our calculations. For the case of the $Z$-\&-$X$-\&-$J$-noise,
we use 40 control parameters in a parameter set to run the two-step
optimization for the two-qubit CNOT gate, and choose 100 initial random
guesses of the parameter sets for the first-step optimization and
10 parameter sets obtained in the first-step optimization as initial
guesses for the second step optimization. We use a total of 60 2GHz-CPU
cores and it takes about 2 days to obtain the control pulses and robust
performance calculations of Fig.~\ref{sFig1}(c). These resources
and time spent to construct the robust high-fidelity CNOT gates against
five sources of high-frequency noise are quite acceptable.

\subsection{Comparison with filter-transfer-function method}

\begin{figure}
\onlinecap \includegraphics[scale=0.60]{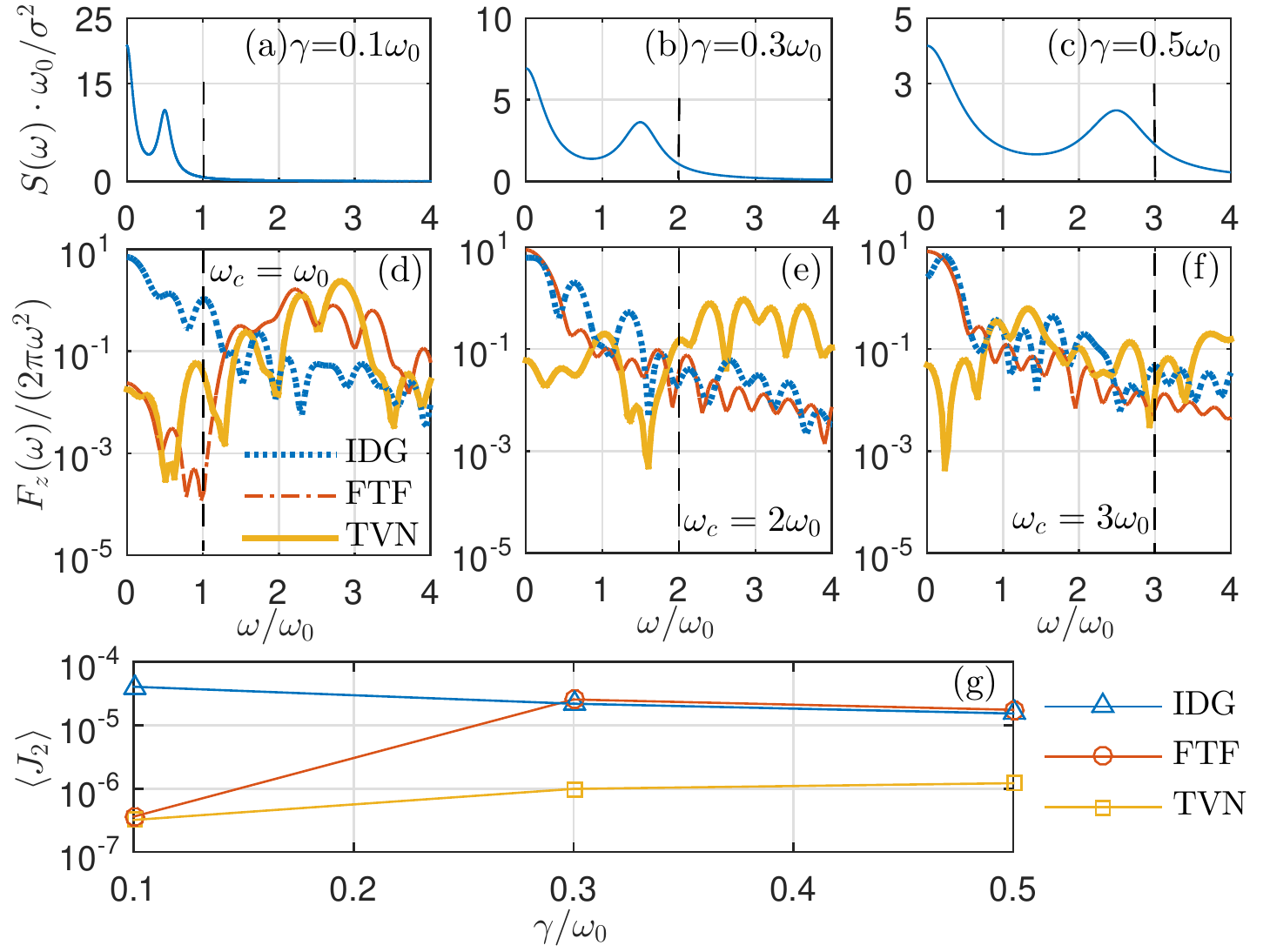}\protect\protect\protect\caption{Behavior of $\left[F_{z}(\omega)/(2\pi\omega^{2})\right]$ obtained
using the optimal control parameter sets from the IDG strategy (thick dotted blue line), FTF (thin dashed orange line)
strategy, and TVN strategy (thick solid yellow line) for the noise PSD $S(\omega)$ with (a)
$\gamma=0.1\omega_{0}$, (b) $\gamma=0.3\omega_{0}$, and (c) $\gamma=0.5\omega_{0}$
are shown in (d), (e), and (f) respectively. Shown in (g) are the
corresponding $\left\langle J_{2}\right\rangle $ values. }

\label{Fig4} 
\end{figure}

In this subsection, we compare our method with the FTF method \cite{SoareBallHayesEtAl2014,BallBiercuk2015,GreenSastrawanUysEtAl2013}.
The cost function $\left\langle J_{2}\right\rangle $ in Eq.\ (\ref{eq:ensemble average j2})
can be transformed to frequency domain as 
\begin{align}
\left\langle J_{2}\right\rangle  & =\underset{j}{\sum}\dfrac{1}{2\pi}\int_{-\infty}^{\infty}\dfrac{d\omega}{\omega^{2}}S_{j}(\omega)F_{j}(\omega),\label{eq:j2_filter_connection}
\end{align}
where $S_{j}(\omega)$ is the noise PSD for the $j$-th noise, and
$F_{j}(\omega)$ is the corresponding filter-transfer function. The
cost function of the $j$-th noise for optimization in the FTF method
is defined as \cite{GreenSastrawanUysEtAl2013,SoareBallHayesEtAl2014,BallBiercuk2015}
$A_{j}\equiv\int_{\omega_{L}}^{\omega_{c}}F_{j}(\omega)d\omega$.
The region $\left[\omega_{L},\omega_{c}\right]$ of the integration
of the cost function $A_{j}$ is determined by the non-negligible
region of the noise PSD. In order to compare with our method, we use
the same form of control pulse, the same number of control parameters,
and the same optimal procedure except changing for the FTF method
the cost function from $J_{1}+\left\langle J_{2}\right\rangle $ to
$J_{1}+A_{j}$ in the second step of the two-step optimization. We
call this procedure the FTF strategy. Then we apply the IDG strategy,
FTF strategy, and TVN strategy to find high-fidelity Hadamard gate
for one-qubit system with single $Z$-noise. To demonstrate the advantage
of our method over the FTF method, we choose the noise PSD for the
$Z$-noise to contain a high-frequency distribution as 
\begin{equation}
S(\omega)=\sigma^{2}\left[\frac{2\gamma}{\gamma^{2}+\omega^{2}}+\frac{\gamma}{\gamma^{2}+(5\gamma-\omega)^{2}}+\frac{\gamma}{\gamma^{2}+(5\gamma+\omega)^{2}}\right]\label{eq:PSD_S}
\end{equation}
that has two peaks at $\omega=0$ and $\omega=5\gamma$. As the value
of $\gamma$ increases, the dominant distribution associated with
the second peak of the PSD $S(\omega)$ moves to a high frequency
region in which the FTF method may not work very effectively. We demonstrate
that our method including the detailed noise PSD distribution in the
cost function can still in this case suppress the gate error coming
from $S(\omega)$, a different PSD from that of the OU noise model
used previously. The lower limit $\omega_{L}$ of the integral of
the cost function $A_{z}$ for the FTF strategy is chosen to be zero,
and the upper limit $\omega_{c}$ is chosen to be $1\omega_{0}$,
$2\omega_{0}$, and $3\omega_{0}$ to enclose the dominant distribution
of $S(\omega)$ {[}see Figs.~\ref{Fig4}(a), (b), and (c){]} for
$\gamma=0.1\omega_{0}$, $0.3\omega_{0}$, and $0.5\omega_{0}$, respectively.
For the single $Z$-noise considered here, the infidelity from Eq.~(\ref{eq:j2_filter_connection})
is $\left\langle J_{2}\right\rangle =\int_{-\infty}^{\infty}d\omega S(\omega)\left[F_{z}(\omega)/(2\pi\omega^{2})\right]$.
The improvement of $\left\langle J_{2}\right\rangle $ can be analyzed
through the overlap of $S(\omega)$ with $\left[F_{z}(\omega)/(2\pi\omega^{2})\right]$
\cite{GordonKurizkiLidar2008}. If the control pulses can make $\left[F_{z}(\omega)/(2\pi\omega^{2})\right]$
small in the dominant distribution region of $S(\omega)$, then $\left\langle J_{2}\right\rangle $
can be significantly improved (reduced). We plot $\left[F_{z}(\omega)/(2\pi\omega^{2})\right]$
evaluated by the optimal control parameter sets obtained from the
above three strategies for three different values of $\gamma=0.1\omega_{0}$,
$0.3\omega_{0}$, and $0.5\omega_{0}$ of $S(\omega)$ in Figs.~\ref{Fig4}(d),
(e), and (f), respectively. The corresponding $\left\langle J_{2}\right\rangle $
values are shown in Fig.~\ref{Fig4}(g). By taking the case of $\gamma=0.3\omega_{0}$
as an example, the function $\left[F_{z}(\omega)/(2\pi\omega^{2})\right]$
of the TVN strategy shows apparent drops near the two peaks of the
noise PSD at $\omega=0$ and $\omega=1.5\omega_{0}$, but the function
for the FTF strategy and the IDG strategy does not. Thus, about one
order of magnitude improvement in $\left\langle J_{2}\right\rangle $
of the TVN strategy over the other two strategies is observed. In
short, as the range of dominant distribution of the PSD enlarges {[}e.g.,
from Fig.~\ref{Fig4}(a) to Fig.~\ref{Fig4}(c){]}, the TVN strategy,
including the detailed noise information (CF) in the optimization
cost function \cite{GordonKurizkiLidar2008}, can suppress the dominant
infidelity contribution more effectively than the FTF strategy. Furthermore,
the concatenation method is used to construct control pulses against
two different non-commuting noises in the FTF method \cite{SoareBallHayesEtAl2014,BallBiercuk2015}.
But using the concatenation method to deal with the case of multi-controls,
multi-sources of noise, and multi-qubits may be very complicated.
On the other hand, our method can find robust control pulses for high-fidelity
CNOT gates that involve three control knobs and up to five sources
of high-frequency noise as demonstrated in Fig.~\ref{sFig1}.

\section{Conclusion}

To conclude, our two-step optimization method can provide robust control
pulses of high-fidelity quantum gates for stochastic time-varying
noise. Besides, our method is quite general, and can be applied to
different system models, noise models, and noise CF's (PSD's). If
the system is very clean, i.e., quantum noise is very weak, our method
presented in this paper can be used directly to find control pulses
for high-fidelity gates in the presence of time-varying classical
noises. For example, decoherence (dephasing) time is rather long in
quantum-dot electron spin qubits in purified silicon
\cite{VeldhorstM.HwangJ.C.YangC.EtAl2014,VeldhorstYangHwangEtAl2015} as
compared to those in GaAs \cite{PettaJohnsonTaylorEtAl2005,KoppensBuizertTielrooijEtAl2006},
that is, quantum noise coming from the coupling to the environmental
spins is very weak, and the dominant gate error is due to the electric
classical control noise \cite{VeldhorstYangHwangEtAl2015}. We will
present the results for quantum-dot electron spin qubits in purified
silicon elsewhere. Our method will make essential steps toward constructing
high-fidelity and robust quantum gates for FTQC in realistic quantum
computing systems. 
\begin{acknowledgments}
We acknowledge support from the the Ministry of Science and Technology
of Taiwan under Grant No.~103-2112-M-002-003-MY3, from the National
Taiwan University under Grant Nos.~105R891402 and 105R104021, and
from the thematic group program of the National Center for Theoretical
Sciences, Taiwan. 
\end{acknowledgments}

\appendix
%dummy comment inserted by tex2lyx to ensure that this paragraph is not empty

\section{Derivation of Eqs. (2)-(4) in the main text}

\label{sec:appendix_a}

We present the derivation of Eqs.~(\ref{eq:expanded infidelity})-(\ref{eq:J2})
in the main text and discuss the role of the extra term $\epsilon$
in Eq.~(\ref{eq:expanded infidelity}). Substituting the total system
propagator in Dyson expansion $U(t_{f})=U_{I}(t_{f})\cdot(I+\Psi_{1}+\Psi_{2}+\cdots)$
into the infidelity definition $\mathcal{I}$ of Eq.~(\ref{eq:infidelity definition})
in the main text, we obtain 
\begin{eqnarray}
 &  & \mathcal{I}=%
%1-\dfrac{1}{4^{n}}\left|{\rm
%    Tr}[U_{T}^{\dagger}U_{I}(t_{f})]\right|^{2}
J_1 \nonumber \\
 &  & \quad-\dfrac{2}{4^{n}}{\rm Re}\{{\rm Tr}[U_{T}^{\dagger}U_{I}(t_{f})]^{\star}\cdot{\rm Tr}[U_{T}^{\dagger}U_{I}(t_{f})\cdot(\Psi_{1}+\Psi_{2}+\cdots)]\}\nonumber \\
 &  & \quad-\dfrac{1}{4^{n}}\left|{\rm Tr}[U_{T}^{\dagger}U_{I}(t_{f})\cdot(\Psi_{1}+\Psi_{2}+\cdots)]\right|^{2}.\label{eq:expanded infidelity-1}
\end{eqnarray}
The first term $J_1$ on the right hand side of the equal sign of Eq.~(\ref{eq:expanded infidelity-1})
is the gate infidelity for the ideal system defined in Eq.~(\ref{eq:J1}). Then we define the error
shift matrix $U_{\epsilon}$ 
of the ideal propagator $U_{I}(t_f)$ at time $t_{f}$
from the target gate $U_{T}$ up to a global phase $\phi$
as 
\begin{equation}
U_{I}(t_{f})=e^{i\phi}U_{T}(I+U_{\epsilon}).\label{eq:ideal propagator}
\end{equation}
Note that when the gate infidelity $J_{1}$ for the ideal system is
made small, the matrix elements of 
$U_{\epsilon}$ also become small.
Substituting the expression of $U_{I}(t_f)$ of 
Eq.~(\ref{eq:ideal propagator}) back to 
%the remaining terms except the first term $J_{1}$ in
Eq.~(\ref{eq:expanded infidelity-1}), we obtain 
\begin{eqnarray}
\mathcal{I} & = & J_{1}+\{-\dfrac{1}{2^{n-1}}{\rm Re}[{\rm Tr}(\Psi_{1})]\}\nonumber \\
% &  & +\{-\dfrac{1}{2^{n-1}}{\rm Re}[{\rm Tr}(\Psi_{2})]-\dfrac{1}{4^{n}}\left|%{\rm Tr}(\Psi_{1})\right|^{2}\}\nonumber \\
 &  & +J_2 +\epsilon(U_{\epsilon},\Psi_{j})+\mathcal{O}(\tilde{\mathcal{H}}_{N}^{m},m\geq3),\label{eq:expanded infidelity-2}
\end{eqnarray}
where $J_2$ is defined in Eq.~(\ref{eq:J2}),
\begin{align}
 & \epsilon(U_{\epsilon},\Psi_{j})=\nonumber \\
 & -\dfrac{1}{2^{n-1}}{\rm Re}\{{\rm Tr}[U_{\epsilon}(\Psi_{1}+\Psi_{2}+\cdots)]\}\nonumber \\
 & -\dfrac{2}{4^{n}}{\rm Re}\{{\rm Tr}[U_{\epsilon}]^{\star}\cdot{\rm Tr}[\Psi_{1}+\Psi_{2}+\cdots]\}\nonumber \\
 & -\dfrac{2}{4^{n}}{\rm Re}\{{\rm Tr}[U_{\epsilon}(\Psi_{1}+\Psi_{2}+\cdots)]\cdot{\rm Tr}\left[\Psi_{1}+\Psi_{2}+\cdots\right]^{\star}\}\nonumber \\
 & -\dfrac{2}{4^{n}}{\rm Re}\{{\rm Tr}[U_{\epsilon}]^{\star}\cdot{\rm Tr}[U_{\epsilon}(\Psi_{1}+\Psi_{2}+\cdots)]\}\nonumber \\
 & -\dfrac{1}{4^{n}}\left|{\rm Tr}[U_{\epsilon}(\Psi_{1}+\Psi_{2}+\cdots)]\right|^{2},\label{eq:extra_term_detail}
\end{align}
and $\mathcal{O}(\tilde{\mathcal{H}}_{N}^{m},m\geq3)$ 
denotes other higher-order terms without containing $U_{\epsilon}$.
The first-order noise term, $-{\rm Re}[{\rm Tr}(\Psi_{1})]/2^{n-1}$,
in Eq.~(\ref{eq:expanded infidelity-2}) actually vanishes. Because the
noise Hamiltonian $\mathcal{H}_{N}$ is Hermitian [with $\beta_{j}(t)$
being real], ${\rm Tr}[\mathcal{H}_{N}(t')]$ is a real number.
Thus the first-order term proportional to the real part of 
${\rm Re}[{\rm Tr}(\Psi_{1})]$,
with ${\rm Tr}(\Psi_{1})=-i\int_{0}^{t_{f}}{\rm Tr}[\mathcal{H}_{N}(t')]dt'$,
 vanishes.
%the real part of ${\rm Tr}(\Psi_{1})$ vanishes. Consequently,
%the first-order term proportional to {\rm Re}[{\rm Tr}(\Psi_{1})]$
%also vanishes.
This result of no first-order noise contribution in $\mathcal{I}$
is similar to that in Ref.\cite{DaemsRuschhauptSugnyEtAl2013}. This
is also the reason why there is no first-order noise contribution in ensemble
average $\langle\mathcal{I}\rangle$ of 
Eq.~(\ref{eq:ensemble average infidelity}). 
%without making the assumption of $\left\langle \beta_{j}(t)\right\rangle =0$.
Equations~(\ref{eq:expanded infidelity})-(\ref{eq:J2}) in the main
text can then be easily obtained from 
Eq.~(\ref{eq:expanded infidelity-2})
with the identification of $\epsilon=\epsilon(U_{\epsilon},\Psi_{j})$.

We discuss below the property and the role of 
$\epsilon=\epsilon(U_{\epsilon},\Psi_{j})$ in  
Eq.~(\ref{eq:expanded infidelity}) or
in Eq.~(\ref{eq:expanded infidelity-2}).
The extra contribution 
$\epsilon\cong\epsilon(U_{\epsilon},\Psi_{j})$
to the gate infidelity with detailed form shown
in Eq.~(\ref{eq:extra_term_detail})
is related to the error shift matrix $U_{\epsilon}$
and all Dyson expansion terms $\Psi_{j}$. As noted earlier, if $J_{1}$ is small,
then the matrix elements of $U_{\epsilon}$ are also small. 
Moreover, if the noise strength
is not too strong such that $|\Psi_{j+1}|\ll|\Psi_{j}|$, then
the extra contribution $\epsilon\cong\epsilon(U_{\epsilon},\Psi_{j})$
is also small.
Therefore when running optimization for small noise strength for which the
higher-order terms $\mathcal{O}(\tilde{\mathcal{H}}_{N}^{m},m\geq3)$
becomes negligible (see Appendix \ref{sec:appendix_b}),
the extra contribution $\epsilon$ can be omitted 
as $J_{1}$ is minimized to a
small number. Consequently,
one can focus on the optimization of
only $J_{1}+\left\langle J_{2}\right\rangle$.

The advantage of introducing $J_1$ and $\epsilon$ in
our method is to enable more degrees of freedom in control parameters 
for optimization. There are actually no $J_{1}$ and the 
extra controbution $\epsilon$ in the gate infidelity
expression of the robust control method of SUPCODE
\cite{WangBishopKestnerEtAl2012,KestnerWangBishopEtAl2013} and
the filter-transfer-function method
\cite{SoareBallHayesEtAl2014,BallBiercuk2015}. 
In these methods, $J_1$ or equivalently 
the error shift matrix $U_{\epsilon}$ is set exactly
to zero by imposing some constraints on the control parameters. 
In contrast, our method can tolerate some error of $U_{\epsilon}$ and
thus have more degrees of freedom 
in control parameters as long as 
$J_{1}$ and the extra contribution $\langle\epsilon\rangle$ in 
gate infidelity $\langle {\cal I}\rangle$ are made just smaller than 
$\langle J_{2} \rangle$.
% without imposing strict
%constraint on the control parameters for making $J_{1}=0$ exactly. 
This advantage of having more degrees of freedom for optimization 
plays an important role in finding better control pulses
 as the number of qubits,
the number of controls, and the number of noise sources increase.

\section{Estimation of
 % how small noise strength is required for neglecting
higher-order contributions
% of $\mathcal{O}(\tilde{\mathcal{H}}_{N}^{m},m\geq3)$
}
\label{sec:appendix_b}

Here we estimate the contributions of higher order terms
$\mathcal{O}(\tilde{\mathcal{H}}_{N}^{m},m\geq3)$ and discuss when they
can be neglected. 
We express the higher-order terms as $\mathcal{O}(\tilde{\mathcal{H}}_{N}^{m},m\geq3)=\sum_{p\geq3}J_{p}$,
where $J_{p}$ denotes the $p$-th order noise term of the gate
infidelity. 
Detailed forms of the first
two lowest-order terms in $\mathcal{O}(\tilde{\mathcal{H}}_{N}^{m},m\geq3)$
are
\begin{align}
J_{3} & =-\dfrac{1}{2^{n-1}}{\rm Re}[{\rm Tr}(\Psi_{3})]-\dfrac{2}{4^{n}}{\rm Re}\{{\rm Tr}(\Psi_{1}){\rm Tr}(\Psi_{2})^{\star}\},\label{eq:J3}\\
J_{4} & =-\dfrac{1}{2^{n-1}}{\rm Re}[{\rm Tr}(\Psi_{4})]-\dfrac{1}{4^{n}}\left|{\rm Tr}(\Psi_{2})\right|^{2}-\dfrac{2}{4^{n}}{\rm Re}\{{\rm Tr}(\Psi_{1}){\rm Tr}(\Psi_{3})^{\star}\},\label{eq:J4}
\end{align}
where
\begin{equation}
  \label{eq:Psi_q}
\Psi_{q}=(-i)^{q}\int_{0}^{t_{f}}dt_{1}\int_{0}^{t_{1}}dt_{2}\cdots\int_{0}^{t_{q-1}}dt_{q}\tilde{\mathcal{H}}_{N}(t_{1})\tilde{\mathcal{H}}_{N}(t_{2})\cdots\tilde{\mathcal{H}}_{N}(t_{q})  
\end{equation}
%$\Psi_{q}=(-i)^{q}\int_{0}^{t_{f}}dt_{1}\int_{0}^{t_{1}}dt_{2}\cdots
%\int_{0}^{t_{q-1}}dt_{q}\tilde{\mathcal{H}}_{N}(t_{1})
%\tilde{\mathcal{H}}_{N}(t_{2})\cdots\tilde{\mathcal{H}}_{N}(t_{q})$  
is the $q$-th order Dyson expansion term.
To make an estimation of the magnitude of $\Psi_{q}$,
we take the 
Z-noise model for the 
single-qubit gate operations in Sec.~\ref{sub:one_qubit}
as an example.
%; and thus $\mathcal{H}_{N}(t)=\beta_{Z}(t)R_{Z}(t)$
%and $R_{Z}(t)=U_{I}^{\dagger}(t)[\omega_{0}Z/2]U_{I}(t)$. We define
%$\bar{R}_{Z}(t)=U_{I}^{\dagger}(t)[Z/2]U_{I}(t)$. 
Substituting the noise Hamiltonian 
$\tilde{\mathcal{H}}_{N}(t)=\beta_{Z}(t)\omega_{0}{R}_{Z}(t)$ 
with $R_{Z}(t)=U_{I}^{\dagger}(t)[\omega_{0}Z/2]U_{I}(t)$
in the interaction picture into $\Psi_{q}$, we obtain 
\begin{align}
\Psi_{q} & =(-i)^{q}\int_{0}^{t_{f}}\omega_{0}dt_{1}\int_{0}^{t_{1}}\omega_{0}dt_{2}\cdots\int_{0}^{t_{q-1}}\omega_{0}dt_{q}\nonumber \\
 & \times\{\beta_{Z}(t_{1})\beta_{Z}(t_{2})\cdots\beta_{Z}(t_{q})\}\{\bar{R}_{Z}(t_{1})\bar{R}_{Z}(t_{2})\cdots\bar{R}_{Z}(t_{q})\}.
\end{align}
where $\bar{R}_{Z}(t)=U_{I}^{\dagger}(t)[Z/2]U_{I}(t)$.
%Next we estimate the magnitude of matrix elements $\left|\Psi_{q,jk}\right|$.
Since $U_{I}(t)$ is unitary, its matrix elements 
$\left|U_{I,jk}(t)\right|\leq1$ for all
$j$ and $k$. Consequently, $\left|\bar{R}_{Z,jk}(t)\right|< 1$ 
for all $j$ and $k$, so is 
$\left|\{\bar{R}_{Z}(t_{1})\bar{R}_{Z}(t_{2})\cdots\bar{R}_{Z}(t_{q})\}_{jk}\right|<1$
for all $j$ and $k$. 
%For stochastic low-frequency OU noise,
%$\beta_{Z}(t)\cong\sigma_{ZZ}$ for most realizations as in Fig.~\ref{Fig1}(c),
%and for high-frequency noise $\beta_{Z}(t)$ evolves up and down with
%initial $\beta_{Z}(t=0)\cong\sigma_{ZZ}$ for most realizations as
%in Fig.~\ref{Fig1}(e). Therefore, 
Taking the strength of $\beta_{Z}(t)$ to be about its standard
deviation $\sigma_{ZZ}$,
we estimate the noise strength
contribution to be
$\left|\{\beta_{Z}(t_{1})\beta_{Z}(t_{2})\cdots\beta_{Z}(t_{q})\}\right|\approx(\sigma_{ZZ})^{q}$.
The time integral contribution
$\{\int_{0}^{t_{f}}\omega_{0}dt_{1}\int_{0}^{t_{1}}\omega_{0}dt_{2}\cdots\int_{0}^{t_{q-1}}\omega_{0}dt_{q}\}$
can be estimated to be about
$\sim (\omega_{0}t_{f})^{q}/q!$. 
By combining the above estimations,  the magnitude of
$\left|\Psi_{q,jk}\right|$ is in the order of
$\sim(\omega_{0}t_{f}\sigma_{ZZ})^{q}/q!$. 
Then substituting the estimated value of
$\left|\Psi_{q,jk}\right|$
%\sim(\omega_{0}t_{f}\sigma_{ZZ})^{q}/q!$ 
into $J_{2}$ in Eq.~(\ref{eq:J2}), $J_{3}$ in Eq.~(\ref{eq:J3}),
and $J_{4}$ in Eq.~(\ref{eq:J4}), 
%we obtain $J_{2}$ has magnitude
%in order of $(\omega_{0}t_{f}\sigma_{ZZ})^{2}/2!$, $J_{3}\sim(\omega_{0}t_{f}\sigma_{ZZ})^{3}/3!$,
%and $J_{4}\sim(\omega_{0}t_{f}\sigma_{ZZ})^{4}/4!$. 
we obtain the magnitude
ratio $J_{3}/J_{2}\sim(\omega_{0}t_{f}\sigma_{ZZ})/3$ and $J_{4}/J_{2}\sim(\omega_{0}t_{f}\sigma_{ZZ})^{2}/12$.
The single-qubit gate operation time in Sec.~\ref{sub:one_qubit} is
$\omega_{0}t_{f}=20$. 
If we choose the noise fluctuation 
$\sigma_{ZZ}=10^{-3}$, then the ratio $J_{3}/J_{2}\sim(6\times10^{-3})$
and $J_{4}/J_{2}\sim(3\times10^{-5})$, and thus the higher-order terms
$\mathcal{O}(\tilde{\mathcal{H}}_{N}^{m},m\geq3)$ can be safely neglected.
If, however, $\sigma_{ZZ}\sim10^{-1}$, then
$\omega_{0}t_{f}\sigma_{ZZ}\sim 2$.
In this case, $J_{3}/J_{2}\sim2/3$ and $J_{4}/J_{2}\sim1/3$, so the higher-order
terms $\mathcal{O}(\tilde{\mathcal{H}}_{N}^{m},m\geq3)$ can not be
neglected.
% Therefore, for our estimation, the order of $\sigma_{ZZ}$
%should be smaller than $\sim10^{-1}$ for neglecting
%$\mathcal{O}(\tilde{\mathcal{H}}_{N}^{m},m\geq3)$.
Comparing our estimation with the results of the
full-Hamiltonian simulation, one finds that the ensemble average of the
gate infidelity $\left\langle \mathcal{I}\right\rangle $
of the IDG strategy scales as the second power of $\sigma_{ZZ}$ (because
$\left\langle J_{2}\right\rangle $ dominates) for small $\sigma_{ZZ}$
until $\sigma_{ZZ}\sim10^{-1}$ in Fig.~\ref{Fig2}(a)
for low-frequency noise $\gamma_{ZZ}=10^{-7}\omega_{0}$ 
and in Fig.~\ref{Fig3}(a) for high-frequency noise
$\gamma_{ZZ}=10^{-1}\omega_{0}$. This is consistent with our estimation.
In other words, if $\sigma_{ZZ}$
is considerably smaller than $10^{-1}$,
$\mathcal{O}(\tilde{\mathcal{H}}_{N}^{m},m\geq3)$ can be ignored. 
%The criterion for neglecting $\mathcal{O}(\tilde{\mathcal{H}}_{N}^{m},m\geq3)$
%by our estimation is equivalent to that by the full-Hamiltonian simulation.
Therefore, even for the case 
when the full-Hamiltonian simulation is not available, we can
use this estimation method to determine the criterion for neglecting
the higher-order terms $\mathcal{O}(\tilde{\mathcal{H}}_{N}^{m},m\geq3)$.

 \bibliographystyle{apsrev4-1}
\bibliography{citation_robust_gates}

\end{document}